\documentclass[journal=langd5,manuscript=article]{achemso}

\usepackage[version=3]{mhchem} 
\usepackage{stix}
\usepackage{xcolor}
\usepackage{siunitx}
\usepackage{graphicx}
\usepackage{caption}
\usepackage{subcaption}
\usepackage{bm}
\usepackage[section]{placeins}
\usepackage{tikz}

\def \rb {\mathbf{r}}
\newcommand{\fig}{Fig.}

\newcommand{\figref}[1]{\fig~\ref{#1}}

\renewcommand{\eqref}[1]{Eq.~(\ref{#1})}

\author{Christopher Kessler}
\affiliation[ITT]
{Institute of Thermodynamics and Thermal Process Engineering, University of Stuttgart, Pfaffenwaldring 9, 70569 Stuttgart, Germany}
\author{Robin Schuldt}
\affiliation[TheoChem]
{Institute for Theoretical Chemistry, University of Stuttgart, Pfaffenwaldring 55, 70569 Stuttgart, Germany}
\author{Sebastian Emmerling}
\affiliation[FKF]
{Nanochemistry Department, Max Planck Institute for Solid State Research, Heisenbergstrasse 1, 70569 Stuttgart, Germany}
\author{Bettina Lotsch}
\affiliation[FKF]
{Nanochemistry Department, Max Planck Institute for Solid State Research, Heisenbergstrasse 1, 70569 Stuttgart, Germany, and Department of Chemistry, University of Munich (LMU), Butenandtstraße 5-13, 81377 München, Germany}
\author{Johannes Kästner}
\affiliation[TheoChem]
{Institute for Theoretical Chemistry, University of Stuttgart, Pfaffenwaldring 55, 70569 Stuttgart, Germany}
\author{Joachim Gross}
\affiliation[ITT]
{Institute of Thermodynamics and Thermal Process Engineering, University of Stuttgart, Pfaffenwaldring 9, 70569 Stuttgart, Germany}
\author{Niels Hansen}
\affiliation[ITT]
{Institute of Thermodynamics and Thermal Process Engineering, University of Stuttgart, Pfaffenwaldring 9, 70569 Stuttgart, Germany}
\email{hansen@itt.uni-stuttgart.de}

\title[An \textsf{achemso} demo]
{Influence of Layer Slipping on Adsorption of Light Gases in Covalent Organic Frameworks: A Combined Experimental and Computational Study}
\abbreviations{GCMC, COFs, COF}

\begin{document}






\begin{abstract}
Sorption of gases in micro- and mesoporous materials \textcolor{black}{is} typically interpreted \textcolor{black}{on the basis of idealized structural models where real structure effects such as defects and disorder are absent}. For covalent organic frameworks (COFs) significant discrepancies between measured and simulated adsorption isotherms are often reported but rarely traced back to their origins. This is because little is known about the real structure \textcolor{black}{of COFs and its effect} on the sorption properties of these materials. In the present work molecular simulations are used to obtain adsorption isotherms of argon, nitrogen, and carbon dioxide in the COF-LZU1 at various temperatures. The (perfect) model COF has a BET surface that is higher than the experimental BET surface by a factor  \textcolor{black}{of approximately 1.33}, suggesting defects or inclusions are present in the real structure. We find that the saturation adsorption loading of small gaseous species in COF-LZU1, as determined from grand canonical Monte Carlo simulations, is also higher by approximately the same factor compared to the experimental saturation loading.  The influence of interlayer slipping on the shape of the adsorption isotherm and the adsorption capacity is studied. Comparison between simulation and experiment at lower loadings suggests the layers to be shifted instead of perfectly eclipsed. The sensitivity of the adsorption isotherms in this regime towards the underlying framework topology shows that real structure effects have significant influence on the gas uptake. Accounting for layer slipping is important to applications such as catalysis, gas storage and separation.

\end{abstract}


\section{Introduction}
Covalent organic frameworks (COFs) are porous materials formed by strong covalent bonds of organic secondary building units (SBUs) composed of light elements (e.g. C, N, H, O)\cite{Cote_2005}. 
They can be classified as two- or three-dimensional polymers in which layers are connected by $\pi$-$\pi$ interactions or by covalent bonds, respectively.
COFs can be characterized by high specific surface areas\textcolor{black}{, crystallinity} and large porosity. In contrast to metal organic frameworks (MOFs) that are also formed by organic linkers but contain metal atoms, the molecular weight of COFs is comparably low\cite{Kushwaha2020}.  These characteristics make COFs intriguing candidates for applications such as gas storage, separation, drug delivery, sensing, catalysis and as optoelectronic materials. \cite{Huang_Angew_2015,Huang_2016,Song_Adv_Sci_2019,Li_IECR_2019,C9MH00856J,Chen_Angew_2020,C9CS00827F,Gottschling_JACS_2020,Nagai_book_2020} The variety of SBUs and the seemingly endless possibilities of their combinations leads to a large number of already synthesized as well as hypothetical frameworks \cite{Lan_2018}. Therefore, it is desirable to choose the "best" COF for the given application and further tune it by changing linkers or incorporating functional groups\cite{mendoza_cortez2011}. Regarding the tremendous amount of possible materials it is necessary to establish COF databases providing structures and properties \cite{TONG2017456,TONG2018,Ongari_ACSCS_2019}. Since carrying out random experiments for such a large variety of possible frameworks without an informed selection is neither feasible nor economic, molecular simulations are a tool of choice to study for example the adsorption properties of COFs. Automated workflows for screening COFs are proposed and applied e.g. for separation, carbon-dioxide capture and other applications \cite{Yan_ACSSCE_2019,Deeg_ACSAMI_2020,Ongari_ACSCS_2020}.

Computer simulations are well established in the field of porous media and led to prection of adsorption and diffusion behavior that are not experimentally explorable. Starting from zeolites and other molecular sieves, great progress considering simulation techniques and force fields has been made \cite{review_zeolites2008,Besley_2022}. Since those materials are ordered crystals, their lattice structures can be experimentally explored by X-ray diffraction, which can be directly transferred to a molecular model. The recent and rapidly emerging research in the field of porous media concentrates on tailored materials such as MOFs and COFs \cite{Barton1999,MOFs_yaghi_rev,Cote_2005}. Due to their similarities we briefly review the new challenges on simulation techniques that both types of materials require and also elaborate on the differences. First publications on adsorption in MOFs already showed promising results regarding the reproducibility of experimental adsorption isotherms by simulation and the usability in fields like energy storage. \cite{Dueren2004,Yang2005,GaberoglioMOFs2005}. For all performed simulations with "ideal", i.e. defect-free crystal structures fully accessible to an adsorbate, it appears natural that adsorption is overestimated,\cite{Ge2018} which has been accounted for by a scaling factor to adjust the excess amount of adsorbate relative to that experimentally observed. \citeauthor{SurbleScale2006} and \citeauthor{DubbeldamScale2007} used the ratio of experimentally obtained pore volume to calculated volume of the perfect crystal of MIL-102 and IRMOF1, respectively, to \textcolor{black}{account for imperfections and thus scale down the simulated loading.} \cite{SurbleScale2006,DubbeldamScale2007}. \citeauthor{Garberoglio_2007} obtained comparable results for argon adsorption in 3D COFs (COF-102 and COF-103).\cite{Garberoglio_2007} Simulations of hydrogen adsorption in 2D (COF-1, COF-5, COF-6, COF-8, COF-10) and 3D (COF-102, COF-103) COFs reported by \citeauthor{AssfourScale2010} used a scaling factor derived by accessible volume.\cite{AssfourScale2010} The crystal structure was obtained by X-ray diffraction without further \textcolor{black}{relaxation by DFT calculations}. The scaled adsorption isotherms for the 3D COFs are in good agreement with the experimental isotherms while for 2D COFs larger deviations occur. Recent experimental studies indicate that structures from X-ray diffraction should be regarded as "average structures" in a crystallographic sense that typically do not take into account real structure effects such as stacking disorder and different stacking polytypes in a single material \cite{Putz_2020}. Although most 2D COFs show random layer offsets to some degree, as demonstrated recently by pair-distribution function analysis and stacking fault simulations \cite{Putz_2020}, the stacking in 2D COFs is simplistically regarded as eclipsed, meaning that the COF-layers are stacked exactly on top of each other. However, assuming ideal eclipsed stacking results in an overestimation of the true adsorption capacity compared to COFs with shifted layers \cite{SharmaSepInterlay2017,Sharma_IECR_2018}. 

The uncertainty on the crystal structure leads to interesting considerations. Adsorption isotherms can be divided into three pressure regimes in which the loading is proportional to the enthalpy of adsorption (low pressure), to the accessible surface area, and to the accessible volume (high pressure)\cite{Forst2006}. All three regimes are more or less influenced by the structural model. We will look at the impact of layer \textcolor{black}{slipping} on various properties in the regimes in the part \textcolor{black}{Study on Interlayer Slipping}. The positions of framework atoms can influence the calculated point charges, as shown by Hamad et al.\cite{Hamad2015} 
The impact of the chosen method to calculate framework charges on adsorption has been extensively studied. For a detailed overview we refer to the literature \cite{Hamad2015, HACKETT2018231, Nazarian2016}. The fact that most zeolites furnish unambiguous crystal structure information and are composed of a limited variety of elements (namely oxygen, aluminium, silicon and counterions) allowed to develop transferable force fields for adsorption and diffusion \cite{Lim_2018,GarciaSanchez2009,Liu2008,Calero2004,Bai2013}. Fitting force field parameters to experimental isotherms in COFs or MOFs is precarious because of the various uncertainties and the fact that error cancellation might lead to non-transferable sets of parameters. Therefore, it is state of the art to use generic force fields such as Dreiding\cite{Dreiding1990} or UFF\cite{RapeUFF1992}. For further details we refer to the literature\cite{DubbeldamReview2019}. 

\textcolor{black}{While deformation and structural transitions of the framework under the influence of adsorbates in MOFs have been simulated using flexible framework models \cite{Schneeman2014,Heinen2018}, such studies are less common for COFs\cite{Amirjalaya2012}. Another challenge concerning MOFs are (unsaturated) metal sites that \textcolor{black}{require} advanced simulation techniques\cite{Chen2011,Zhao2021}.}
Despite the existence of databases, screening studies and investigations of adsorption in COFs,
a quantitative prediction of isotherms is still challenging.
With the presented techniques and state of the art methodology it is in general possible to get a rather qualitative view of the adsorption isotherm, which is often sufficient to reduce the number of suitable COFs for further detailed experimental or computational investigation of the properties.

In this work we combine experimental studies on the imine-linked 2D COF-LZU1\cite{DingCOFLZU12011} with computational methods to elucidate the impact of the various degrees of freedom that need to be fixed when defining a computational model. By studying the influence of adsorbate-framework interaction strength and layer \textcolor{black}{slipping} on the adsorption properties of small gaseous species,     
\textcolor{black}{\textcolor{black}{real structure effects} can be partially separated from uncertainties in the computational model, thus allowing to highlight the need for improved strucutural models of covalent organic \textcolor{black}{frameworks}.}

\section{Experimental Work}
For the synthesis of COF-LZU1 a modified room temperature synthesis method developed by Shiraki et al. was used.\cite{Shiraki2015synthesis} In a glass vial 1,3,5-triformylbenzene (49.9~mg, 0.308~mmol) and \emph{para}-phenylenediamine (50.0~mg, 0.462~mmol) were fully dissolved in 5~mL 1,4-dioxane and 0.5~mL 3~M aqueous acetic acid was added. The vial was closed and after sonicating the mixture for one minute, it was left standing at room temperature for 48~h. The resulting solid was filtered off, washed with Aceton, DMF, and methanol and then subjected to Soxhlet extraction with MeOH for 16~h. The MeOH soaked solid was then activated by supercritical CO\textsubscript{2} drying and further under high vacuum for 24~h to obtain 77.2~mg COF-LZU1 (93~\%). Fourier-transform infrared spectroscopy (FTIR) and X-Ray powder diffraction (XRPD) analysis showed a successful COF formation in accordance with literature, see also Figures S1 and S2.

The sorption measurements for this work were performed on a Quantachrome Instruments Autosorb iQ~3 with \textcolor{black}{ argon as adsorbate at 87~K, nitrogen at 77~K and 273~K as well as CO$_2$ at 273~K, 288~K and 298~K}. Before measurement, the sample was activated under high vacuum at 120~°C for 12~h. The experimental pore size distribution was determined from the nitrogen adsorption isotherm using the NLDFT (cylindrical pores, adsorption branch) kernel for nitrogen at 77~K implemented in the ASiQwin software v~3.01. 

\section{Computational Studies}
\subsection{Computational Details of the DFT-Calculations}
\subsubsection{Cell Optimization}
The crystal structure of the utilized COFs was extracted from X-ray powder diffraction. \textcolor{black}{Comparison to the results of \citeauthor{DingCOFLZU12011} shows good agreement. \cite{DingCOFLZU12011}} 
The experimentally obtained initial structures were optimized using Kohn-Sham density functional theory (DFT) with the Gaussian and plane-waves (GPW) approach as implemented in the quickstep module \cite{lippert1997hybrid,vandevondele2005quickstep} within the CP2K 5.1 program package \cite{kuhne2020cp2k}. Cell optimizations with simultaneous geometry optimization were performed, under constrained hexagonal cell symmetry, applying the LBFGs algorithm. For the exchange and correlation functional we used the PBE generalized gradient correction \cite{perdew1996generalized}, with Grimme's D3 dispersion correction \cite{grimme2010consistent,grimme2011effect}. For all framework atoms, TZV2P-GTH-PBE \cite{vandevondele2007gaussian} basis sets were used. 
For all atoms, corresponding GTH-PBE pseudo potentials \cite{goedecker1996separable} were used for the non-valence electrons.
During the optimization, we used a plane wave cutoff of 300 Ry with an $\mathrm{EPS\_DEFAULT}$ value of $10^{-12}$ determining the cutoff for considered overlap elements. The self consistent field (SCF) calculations were considered converged after reaching the target accuracy of $10^{-6}$ Hartree. Optimizations were performed for two layered unit cells of the investigated COF-LZU1, with systems created based on two distinct configurations: one corresponding to perfectly eclipsed stacking and one with layers shifted with respect to each other. The shifted initial structure was created by displacing the top layer by 1.55 \AA \space along the $x$-direction from the eclipsed configuration. 

\subsubsection{Charge Calculations}
Atomic partial charges were calculated based on the electronic density obtained from the optimized unit cells. The density was calculated based on single-point calculations using a plane-wave cutoff of 500 Ry, an $\mathrm{EPS\_DEFAULT}$ value of $10^{-14}$ and a SCF target accuracy of $10^{-7}$ Hartree. The obtained density distribution was extracted and used together with the optimized unit cell for the calculation of DDEC6 partial charges \cite{manz2016introducing,manz2017introducing,limas2016introducing,limas2018introducing}, as implemented in the Chargemole 3.5 program package.

\subsection{Random Path Generation}
\label{sec.FRDG}
\begin{figure}[ht!]
\begin{subfigure}{0.45\linewidth}
    \centering
    \includegraphics[height=0.6\linewidth]{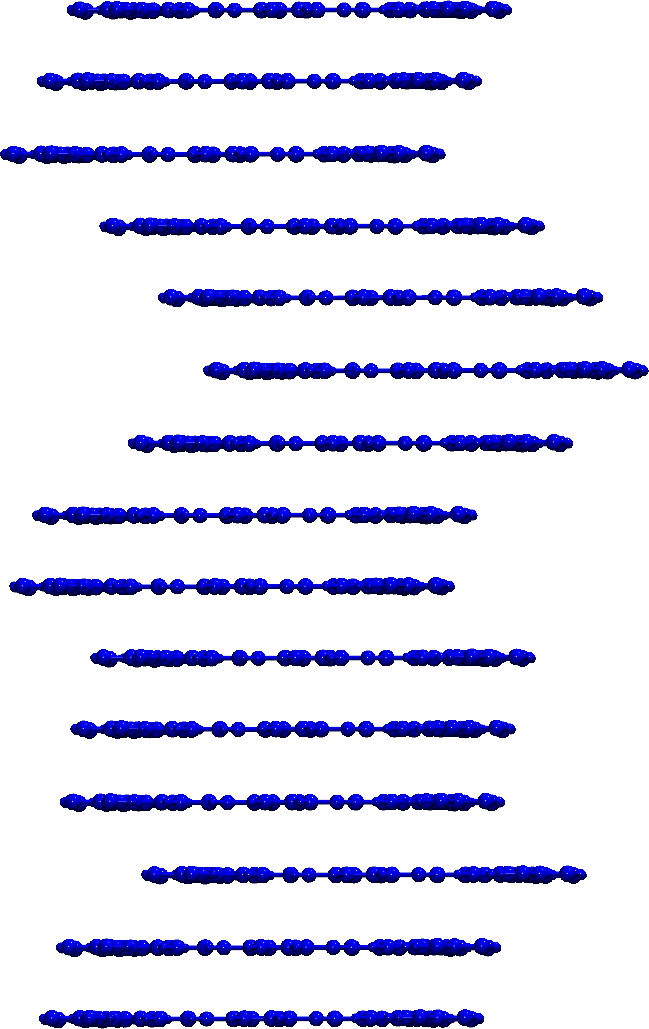}
    \caption{\quad}
    \label{fig.randompath-1}
\end{subfigure}
\begin{subfigure}{0.45\linewidth}
    \centering
    \includegraphics[height=0.6\linewidth]{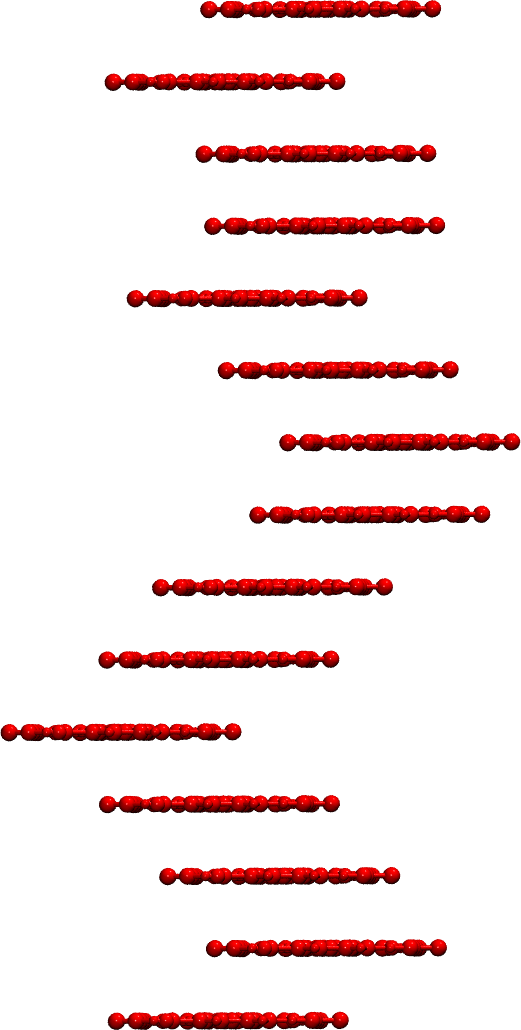}
    \caption{\quad}
    \label{fig.randompath-2}
\end{subfigure}
    \caption{Example of a generated random path for a target in-plane displacement of $d = 5$~\r{A} for COF-LZU1, and an interlayer distance of $3.601$ \r{A} in stacking direction, in agreement with the experiment. (a) The view aligned with the $x$-axis. (b) The view aligned with the $y$-axis.}
    \label{fig.randompath-overview}
\end{figure}

For the generation of random stacking schemes, an in-house code was utilized, creating random displacements in the $x-y$ plane for every layer stacked in $z$-direction. The random paths were generated by duplicating one layer and displacing it by 3.601 \r{A} in $z$-direction, in agreement with the experimentally obtained interlayer distance \cite{DingCOFLZU12011}. The stacked layer is then further displaced by the random shift in the $x-y$ plane. Displacements associated with an individual layer were chosen in the range of 1.55 to 5.0~\r{A}\textcolor{black}{, which includes the one observed in DFT-optimized cells of approximately 2.19~\AA,} in order to gain an overview of their impact on the obtained isotherms.  The created paths were constructed under the constraint of avoiding large in-plane displacements over the cell boundary in stacking direction. As starting point, paths based on an initial guess of random displacements were created. In order to obtain a periodic displacement function, these were obtained by discrete inverse Fourier transform of, randomly chosen, complex coefficients. The obtained displacement lengths were then optimized by a spring model to match the desired  displacement lengths (see Supporting Information for details).
An examplary path generated with the explained algorithm is shown in \figref{fig.randompath-overview}. Acting as an example, the chosen path was constructed for the largest used displacement of 5~\r{A}, to improve visibility of the shifts. The displacement paths in $x$ and $y$ direction are illustrated in \figref{fig.randompath-1} and \figref{fig.randompath-2} by aligning the view with the $y$- and $x$-axis, respectively.
\FloatBarrier
\subsection{Grand Canonical Monte Carlo (GCMC) Simulation}
Adsorption isotherms were obtained by the GCMC methodology\cite{Norman_1969} using the molecular simulation software package RASPA \cite{Dubbeldam2016}. \textcolor{black}{Figures visualizing the pore topology were created with iRASPA\cite{Dubbeldam2018}.} To describe adsorbate-adsorbate interactions the TraPPE force field\cite{Potoff2004} was used for nitrogen and carbon-dioxide and Lennard--Jones parameters by \citeauthor{Michels1949} for argon \cite{Michels1949}. The quadrupole moment of nitrogen was taken into account by placing negative point charges (-0.482~e) on each nitrogen atom which is neutralized by a positive point charge (0.964~e) in the center of mass of the molecule. Carbon dioxide is a three site model with a point charge on each atom and argon consists of a single uncharged Lennard–Jones bead. 
The COF framework was considered to be rigid such that only Lennard--Jones parameters and partial atomic charges needed to be assigned to the different atomic species. 
The Lennard–Jones parameters were taken from the DREIDING force field.\cite{Dreiding1990}
Parameters for unlike Lennard–Jones interaction sites were determined using Lorentz–Berthelot 
combining rules.\cite{Lorentz1881, Berthelot1898} 
The real part of the electrostatic interactions was evaluated up to a cut-off radius of  12.0 \AA. Long-range electrostatic interactions were calculated by Ewald summation\cite{Ewald1921, Dubbeldam2013} with a relative precision of $10^{-6}$. We compute van der Waals interactions up to a spherical cutoff radius of 14 \AA.
Although the density beyond the cut-off radius is not uniform, we applied analytic corrections to the long-range Lennard–Jones tail, in order to reduce the sensitivity of the results with respect to the cut-off radius\cite{Jablonka_2019}. All force field parameters used in the present work are reported in the Supporting Information. 

The simulated state points are chosen to match the experimental adsorption isotherms. In GCMC simulations the pore is in equilibrium to a bulk phase which is represented by the chemical potential and temperature. The pressure in experiments is converted to fugacity, i.e. the chemical potential, using Peng–Robinson equation of state (EOS). The applicability of Peng–Robinson–EOS for the given adsorbates and state points was checked by comparing the results to the ones of PC-SAFT EOS\cite{GrossSadowski2001}. As experimental data does not show significant hysteresis we do not expect it to occur in simulation and so simulated all points in parallel. If one would be looking for hysteresis, we had to use a simulation at high pressure as starting condition and subsequently decrease pressure to follow the desorption branch.

The chosen Monte Carlo moves were insertion, deletion, translation, and rotation. The number of MC cycles was at least $25\times10^3$, both, for equilibration and for the production phase. One cycle consists of $\max(20, N_t)$ MC moves (with $N_t$ as the sum of adsorbate molecules in the system).\\
The observable in GCMC simulations is the absolute number of molecules in the simulation box, while 
experimentally the excess adsorption is measured. The relation between the two quantities is
\begin{equation}
  n_{\mathrm{excess}} = n_{\mathrm{absolute}} - \rho_{\mathrm{bulk}} V_{\mathrm{pore}} \label{eqn:excess}
\end{equation}
where $V_{\mathrm{pore}}$ is the free pore volume and $\rho_{\mathrm{bulk}}$ the density of the equilibrium bulk phase calculated by Peng–Robinson EOS.\cite{Talu2001}
In simulation we compute the pore volume by random insertions of a helium probe atom at a temperature of 298 K in the framework and subsequent evaluation of the Boltzmann factor according to \cite{ref_pot_talu2001}
\begin{equation}
   V_{\mathrm{pore}} = \int e^{-\frac{\phi_i\left( \rb \right)}{k_B T}} \mathrm{d} \rb\label{eqn:void_kb}
\end{equation}
where \textcolor{black}{$\phi_i = \sum_j \phi_{ij}$ is the potential energy between a test molecule of type $i$ and all solid (COF) interaction sites $j$. Further $\phi_{ij}$ is the pair potential.} Note that eq. (\ref{eqn:void_kb}) is not simply a geometric evaluation, \textcolor{black}{the results depend on the choice of the test species $i$, which is helium in our case, through the resulting} $\epsilon_{ij}$- and $\sigma_{ij}$- Lennard–Jones parameters and on temperature. \textcolor{black}{The free volume calculated here is based on an ideal COF structure, which does not take into account cracks or powder packing occurring in experiments.} \textcolor{black}{Pore size distribution and surface area} are (in contrast to eq. (\ref{eqn:void_kb})) calculated by purely geometric evaluation i.e. they are only dependent on $\sigma_{ij}$ of the test molecule and solid atoms. To do a geometric evaluation the test molecule is inserted randomly in a certain distance to the framework atom and then checked for overlaps to framework atoms. Common choices for the distance are $\sigma_{ij}$ or $2^{1/6} \sigma_{ij}$ which is the well-depth of the Lennard–Jones potential. The latter yields smaller values for surface area and pore size distribution is shifted to smaller values. In the present work we chose $\sigma_{ij}$ of hydrogen. The cumulative pore volume function $V_{\mathrm{pore}}\left( r \right)$ is obtained by searching for the largest sphere that overlaps \textcolor{black}{grid points} in the \textcolor{black}{free volume of the} simulation box without overlapping any framework atoms\cite{Sarkisov2011,Gelb1999}. The  pore size distribution, which can be obtained experimentally by applying an isotherm model, is the derivative of the cumulative pore volume function $\frac{\mathrm{d} V_{\mathrm{pore}}\left(r\right)}{\mathrm{d}r}$.

\section{Results and Discussion}
\subsection{COF Structure}
\label{sec.COF-structure}
\begin{figure}
\begin{subfigure}{0.45\linewidth}
    \centering
    \includegraphics[height=0.5\linewidth]{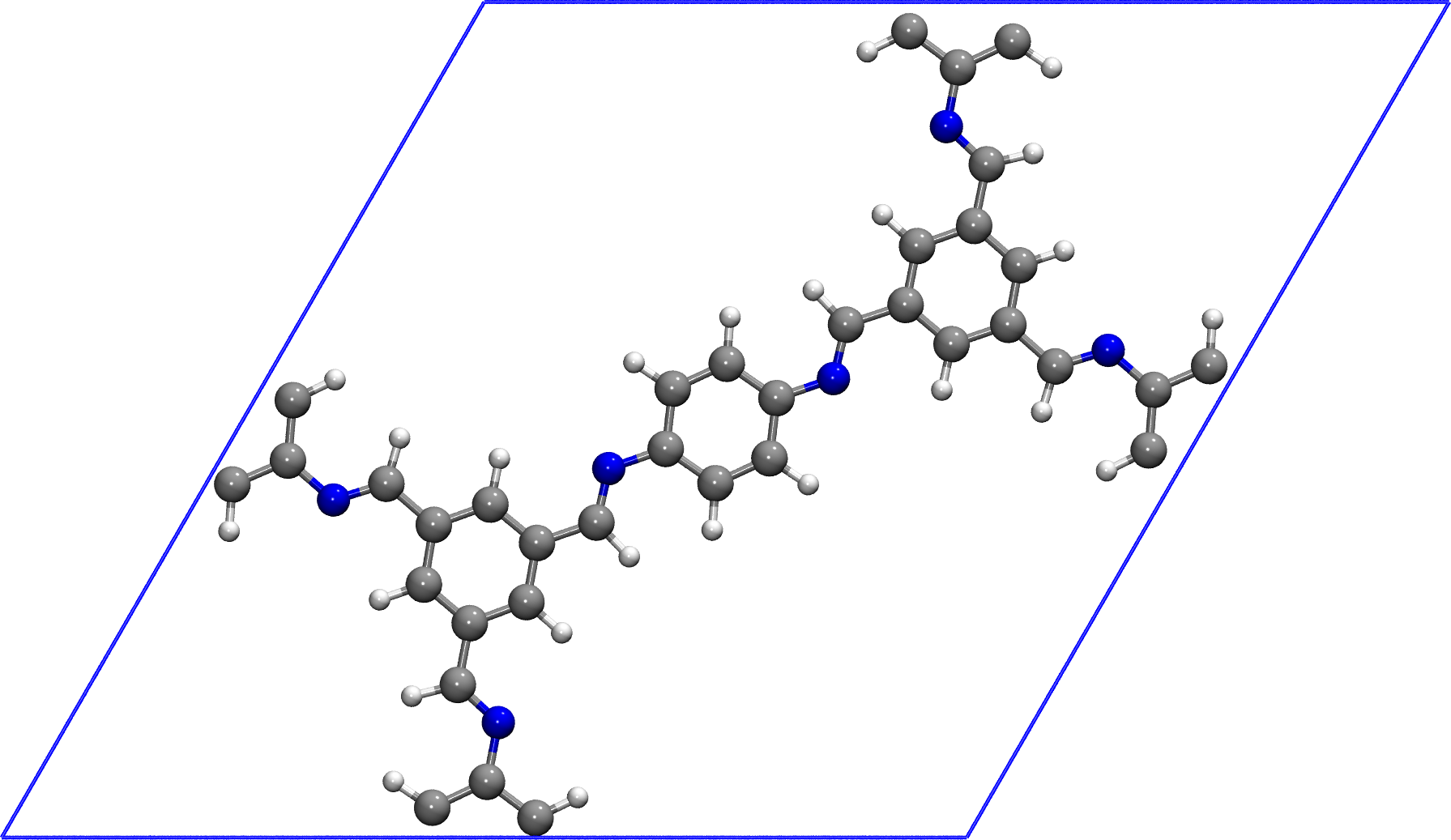}
    \caption{\quad}
    \label{fig:N-COF-unitcell}
\end{subfigure}
\begin{subfigure}{0.45\linewidth}
    \centering
    \includegraphics[height=0.5\linewidth]{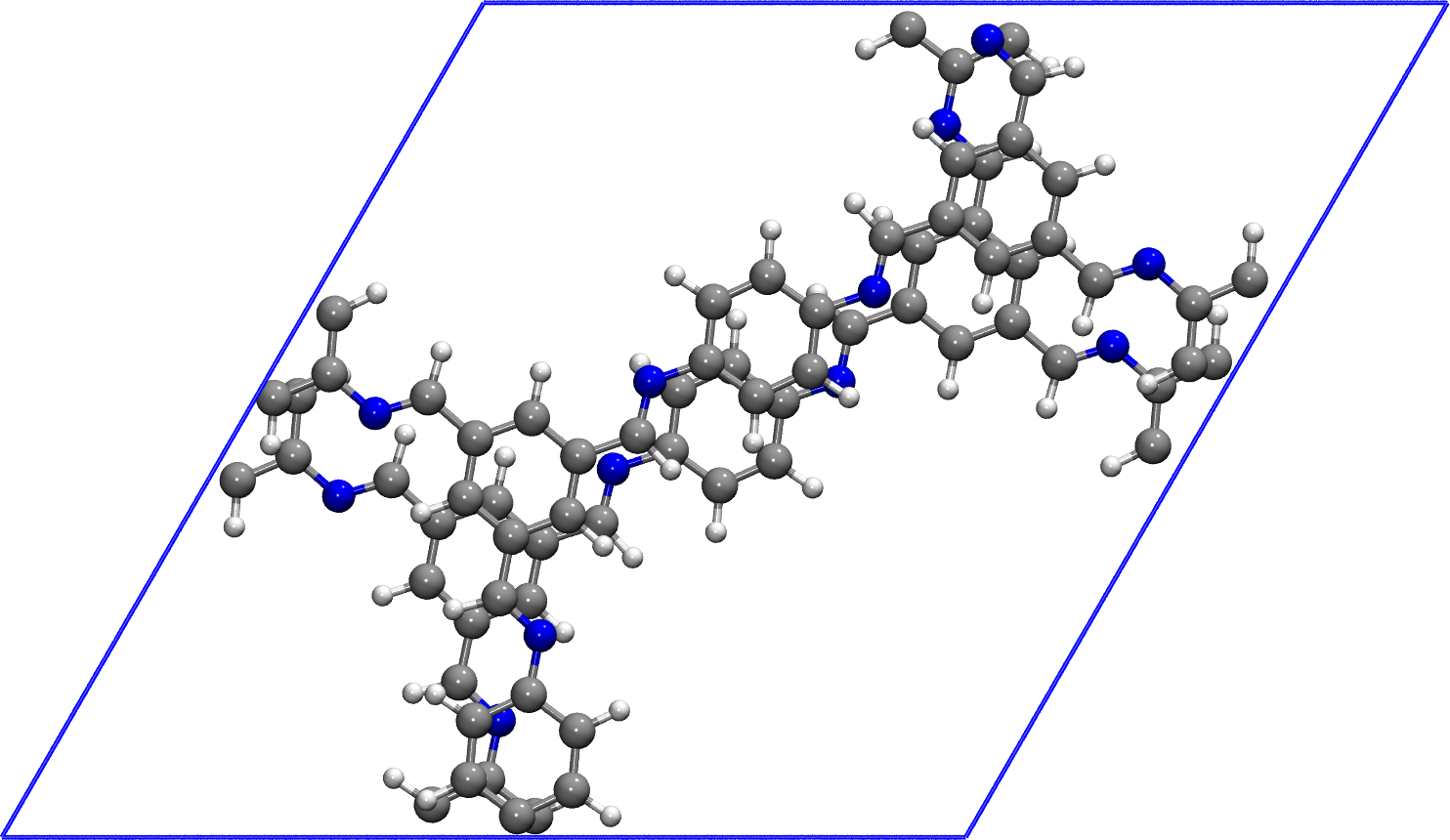}
    \caption{\quad}
    \label{fig:N-COF-shift-unitcell}
\end{subfigure}
    \caption{Top view onto the unit cell of the investigated COF-LZU1, with the unit-cell boundaries in the $x$-$y$-plane indicated in blue. (a) Shows the optimized unit cell for the eclipsed configuration whereas in (b), the shifted structures are illustrated.}
    \label{fig:N-COF-overview}
\end{figure}

The DFT-optimized cells used in the following simulations are shown in \figref{fig:N-COF-overview}. A comparison of the total energies calculated by means of DFT, showed that slightly displaced layers are energetically more favourable by $\Delta E \approx 15.32$~kcal~mol$^{-1}$. The obtained structures after cell optimization are planar without any notable corrugation in both cases. The cell parameters of the hexagonal unit cell were obtained as $a = b = 22.470$~\r{A} and $c = 8.101$~\r{A} for the eclipsed initial configuration shown in \figref{fig:N-COF-unitcell}. For the shifted initial structure, the cell parameters after optimization were obtained as $a = b = 22.465$~\r{A} and $c = 7.274$~\r{A}, the resulting structure is illustrated in \figref{fig:N-COF-shift-unitcell}.

\subsection{Partial Atomic Charges}

\begin{figure}[htb!]
    \centering
    \includegraphics[width=0.5\textwidth]{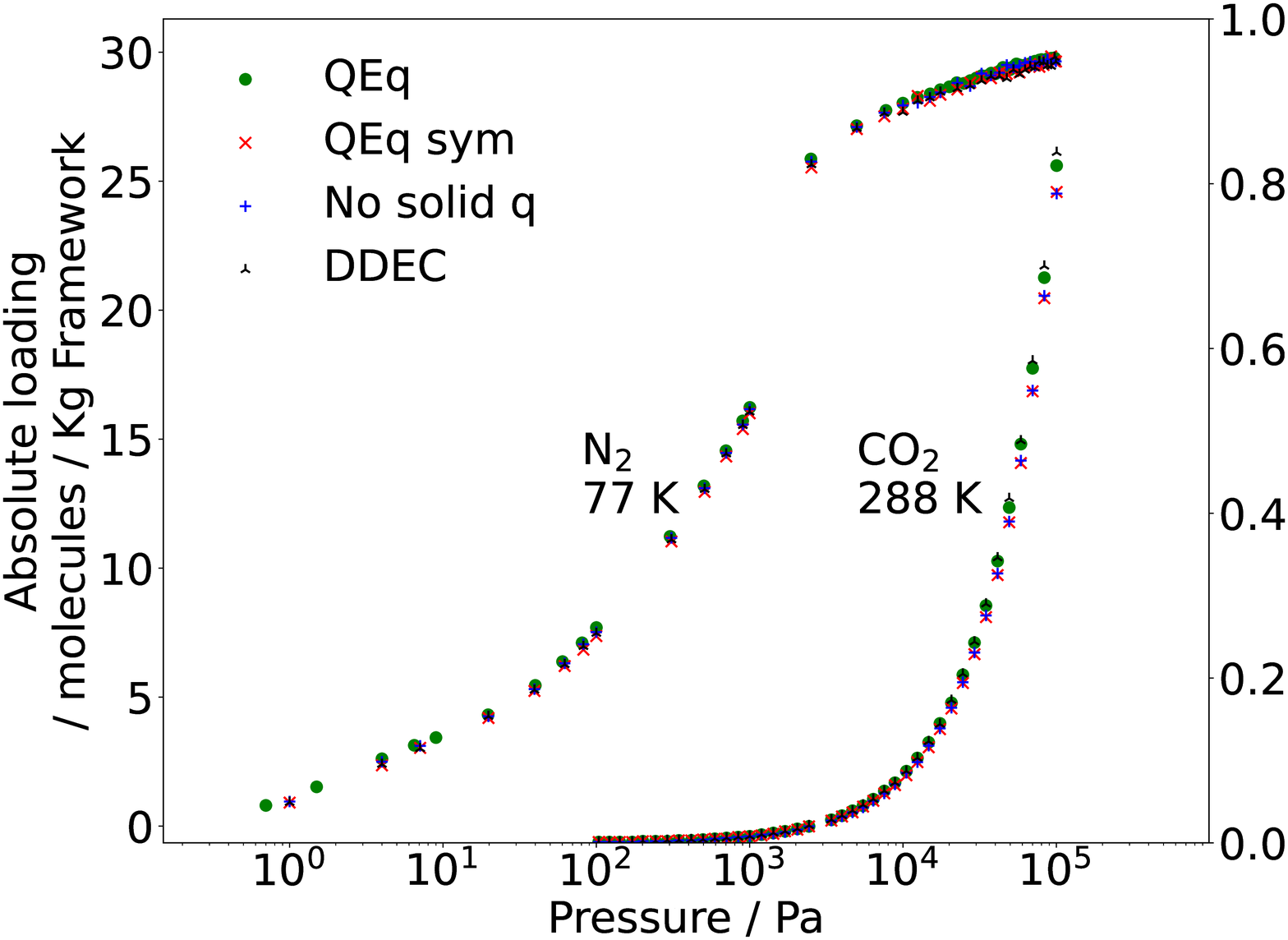}
   \caption{Pure component adsorption of nitrogen at 77 K and carbon dioxide adsorption in the shifted structure at 288 K with various point charges of the solid atoms. CO$_2$ is displayed on the secondary $y$-axis.}
    \label{fgr:charges}
\end{figure}

To analyze the influence of point charges of the framework atoms on the adsorption behavior, we simulated nitrogen adsorption at 77 K and carbon-dioxide adsorption at 288 K in the DFT-optimized structure. For comparison, we used symmetrized DDEC6 charges, extended charge equilibration (EQeq)\cite{Willmer2012} charges with and without symmetrization and neglected charges of the solid atoms. The results are plotted in Figure \ref{fgr:charges}. The influence of the point charges on nitrogen adsorption in COF-LZU1 is nearly insignificant. For carbon dioxide the influence at 288 K is larger, so that neglecting solid charges leads to an error up to 10 \%. It has been shown by \citeauthor{HACKETT2018231}\cite{HACKETT2018231} for sileceous zeolites that the influence of partial atomic charges on the adsorbed loading is strongly dependent on the framework geometry and the magnitude of adsorbate-framework charges. 
Comparison of several charge-assignment schemes for $\mathrm{CO}_2$-adsorption in MOFs also showed a significant effect in many cases.\cite{sladekova2020,sladekova2021} In recent work, we showed that the influence of partial charges for COF TpPA-1 and COF 2,3-DhaTph, which are quite similar in pore size and atomic structure, is of minor importance for nitrogen adsorption at higher temperatures\cite{GCMC_DFT}. 
To estimate the impact of using a non-polarizable force field on the adsorption isotherm, we recalculated partial atomic charges with the DDEC6 approach for COF structures loaded with adsorbate molecules. These new charges were then used in a subsequent GCMC simulation at the same pressure. However, the changes in the amount adsorbed were found to be not significant, leading us to conclude that framework partial charges play a minor role\textcolor{black}{, compared to other influences, for this COF system and for the considered adsorbates.}

\subsection{Scaling factor}
\begin{figure}[htb]
    \centering
    \includegraphics[width=0.5\textwidth]{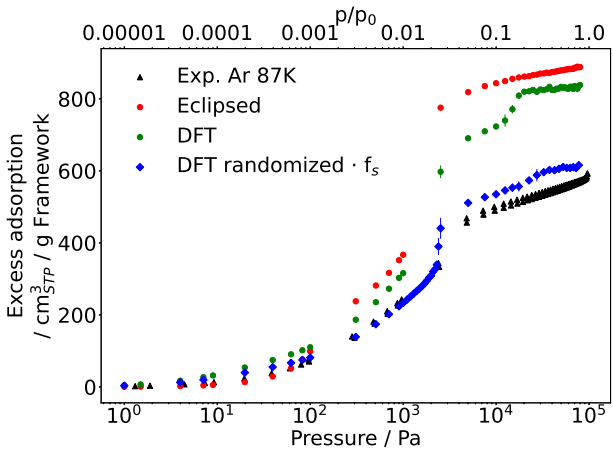}
    \caption{Pure component adsorption of argon at 87~K. Red symbols represent loading in the perfectly eclipsed structure (see Fig.~\ref{fig:N-COF-unitcell}). Green circles represent adsorption in a DFT-optimized version of the framework (see Fig.~\ref{fig:N-COF-shift-unitcell}). Scaled results are displayed by blue diamonds in structure shifted by 2.19 \AA, which is the shifting distance observed in DFT-optimization but in a random direction which removes the step at $p \approx 2\cdot 10^4$~Pa. Black symbols show the experimental adsorption isotherm.}
    \label{fgr:ar_87K}
\end{figure}

\begin{table}
  \caption{Brunauer–Emmett–Teller (BET) surface areas\cite{Brunauer1938} from experimental and simulated adsorption isotherms}
  \label{tbl:materialproperties}
  \begin{tabular}{llll}
    \hline
     & Experiment & Simulation DFT opt.& $f_\text{S}$\\
    \hline
    S-BET$_{\text{Ar}}$ / m$^2$/g   & 1973.7 & 2581.9  & 0.76 \\
    S-BET$_{\text{N}_2}$ / m$^2$/g  & 1815.3 & 2506.3  & 0.72 \\
    S-Connolly / m$^2$/g           &        & 2468.0  &      \\
    \hline
  \end{tabular}
\end{table}
To compare experimental adsorption isotherms in MOFs with simulated ones, Becker et al. \cite{Becker2017} used a scaling factor of 0.85 to account for imperfections and blocked pores in the real structure. Likewise, Assfour and Seifert applied a scaling factor to reach agreement between simulated and experimental adsorption isotherms of hydrogen in COFs.\cite{AssfourScale2010}
In the present study, we also notice that simulated adsorption tends to be higher than experiment. Therefore, we use a scaling factor $f_{\text{s}}$ which is related to the surface area:
\begin{equation}
  f_{\text{s}} = \frac{S_{\text{BET,exp}}}{S_{\text{BET,sim}}} \label{eqn:scalingfactor}
\end{equation}
Assuming circular pores this factor not only scales the accessible surface area but also the pore volume. 
\begin{figure}[htb]
    \centering
    \includegraphics[width=0.5\textwidth]{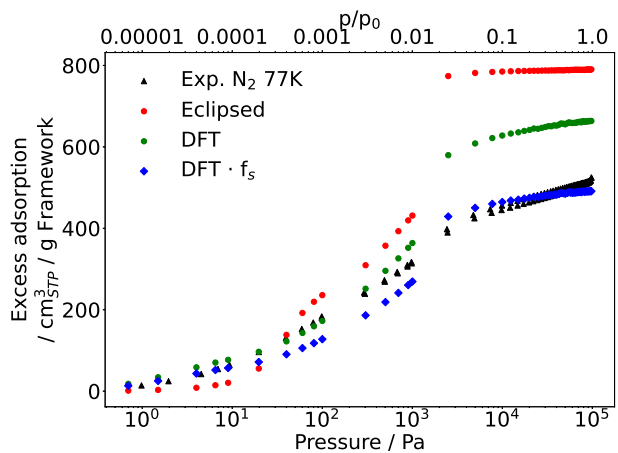}
    \caption{Pure component adsorption of nitrogen at 77~K. Red symbols represent loading in the perfectly eclipsed structure (see Fig.~\ref{fig:N-COF-unitcell}). Green circles represent adsorption in a DFT-optimized (i.e. shifted) version of the framework (see Fig.~\ref{fig:N-COF-shift-unitcell}). Scaled results are displayed by blue diamonds, while black symbols show the experimental adsorption isotherm.}
    \label{fgr:n2_77K}
\end{figure}
When deciding whether to use volume or surface to determine a scaling factor one has to consider that porosity determined from adsorption isotherms uses several assumptions such as round pores that have to be verified experimentally to get a physical basis. Also BET theory relies on several assumptions that have to be checked when applying it\cite{Walton2007}. We decided that the most convincing method to calculate the scaling factor is to determine BET surfaces from experimental isotherms and simulated ones. The calculation was performed using the Python package pyGAPS\cite{pyGAPS2019}. We also chose to utilize the adsorption isotherm of the DFT-optimized structure of COF-LZU1 for calculating the BET surface. Results are given in Table \ref{tbl:materialproperties}. Following the procedure for argon and nitrogen we obtained scaling factors of 0.76 and 0.72, respectively. As both factors are similar we chose to use their mean, 0.74, for further calculations. The Connolly surface area \cite{Dueren2007, Connolly1983} of the framework, a purely geometric calculation, yields a scaling factor in the same range \textcolor{black}{(argon 0.80, nitrogen 0.73)}.
To discuss the necessity of scaling factors we analyze adsorption at lower temperatures. Figure \ref{fgr:n2_77K} shows a comparison between experiment and simulation for nitrogen adsorption at 77~K. The adsorption isotherm of the eclipsed structure distinctly overestimates adsorption even if one would apply the scaling factor. Simulated adsorption in the DFT-optimized (shifted) structure reproduces the experimental adsorption isotherm qualitatively. 
\textcolor{black}{ It should be noted here that the lower saturation capacity of the shifted structure is mainly caused by the 10\% decrease of the unit cell size in $z$-direction compared to the eclipsed structure during DFT optimization.}
At low pressures, the simulation results are seen in good agreement with experimental results. Using the described scaling law we can reproduce the saturation loading quantitatively but underestimate loading at lower pressures.
Considering argon adsorption at 87~K displayed in Figure \ref{fgr:ar_87K}, we found similar results.
In contrast to nitrogen, a step is observed in the isotherm 
at $p \approx 2\cdot 10^4$~Pa calculated in the DFT-optimized shifted structure.
This could be traced back to the regular zig-zag structure of the layers such 
that the size of argon is commensurate with the space between two next-nearest layers.
By keeping the same shifting distance from layer to layer but generating a random path the \textcolor{black}{behavior} disappears and saturation loading is only marginally overestimated after scaling. For argon we find that scaling does not lead to underestimations in the low pressure regime. 
However, as shown in the next section in the low pressure regime, the amount adsorbed is very sensitive to small changes in the strength of the interaction 
between fluid and solid.
For further analysis on interlayer slipping we generally apply the scaling factor of 0.74 for nitrogen at 77~K and argon at 87~K.


\FloatBarrier
\subsection{Force field}
\begin{figure}[htb]
    \centering
    \includegraphics[width=0.5\textwidth]{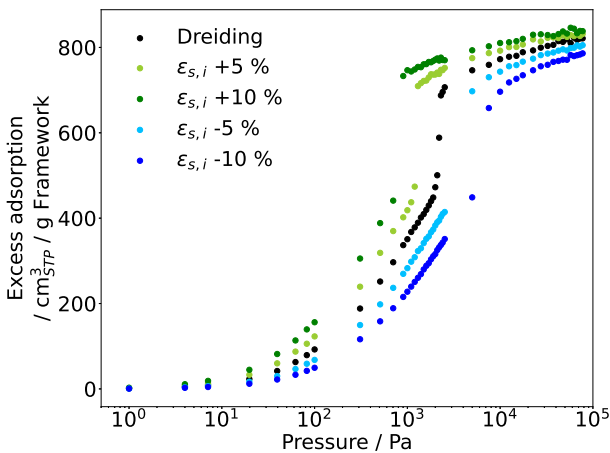}
   \caption{Pure component adsorption of argon at 87.3 K for the randomly shifted structure. The Lennard–Jones energy-parameters of the solid atoms are varied.}
    \label{fgr:forcefield_var}
\end{figure}

The force fields used for the gas molecules have been calibrated to 
reproduce vapor-liquid equilibrium, while the DREIDING force field is commonly 
used to model the framework atoms in COFs. The impact of a variation of the 
Lennard–Jones energy parameter of the framework atoms by up to 10\% on low-temperature argon adsorption isotherms is illustrated in Figure \ref{fgr:forcefield_var}.
It can be observed that varying the force field parameters most prominently affects the low-pressure regime \textcolor{black}{($p <2\cdot10^3$~Pa), where} adsorption is governed by the solid-fluid interactions \textcolor{black}{and available surface area}. 
\textcolor{black}{At higher pressures the adsorption behavior is mainly determined by the free volume, so that fluid-fluid interactions are controlling the amount adsorbed and solid-fluid Lennard–Jones parameters are of minor importance.}
It can therefore be concluded that the underestimation of scaled nitrogen adsorption isotherm discussed above could be remedied 
by slightly increasing the fluid-solid interaction strength.
\FloatBarrier
\subsection{Study on interlayer slipping}\label{ssec:interlayer}
\begin{figure}[htb]
    \centering
    \includegraphics[width=0.5\textwidth]{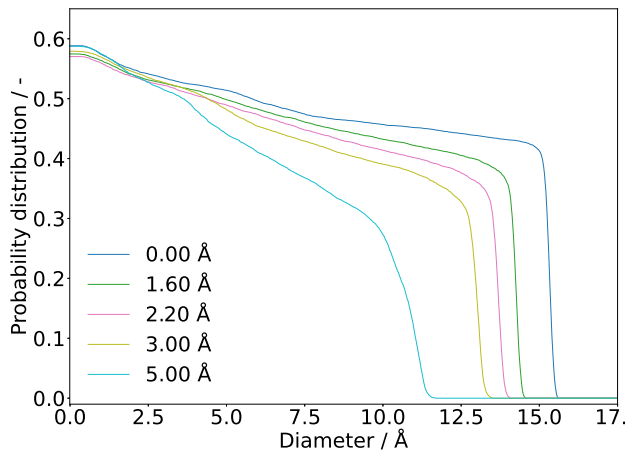}
    \caption{Cumulative pore volume curves \textcolor{black}{(obtained by geometric evaluation)} in different random path COF structures. Here, the shifting of 2.2~\AA~is almost the same as that found in the DFT-optimized structure.}
    \label{fgr:prob_dist}
\end{figure}

\begin{figure}[htb]
    \centering
    \includegraphics[width=0.5\textwidth]{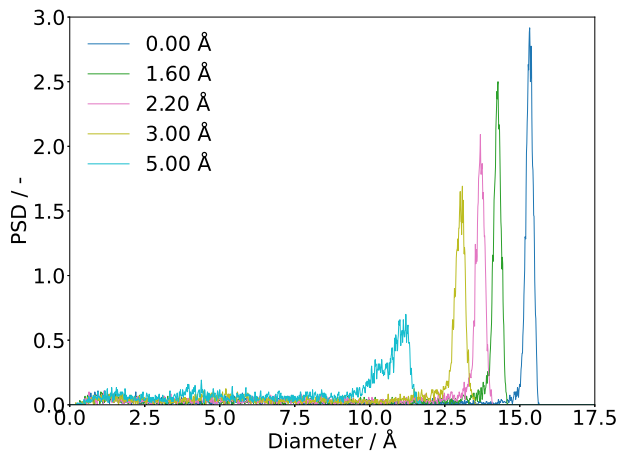}
    \caption{Pore size distributions (PSD) in different random path COF structures.}
    \label{fgr:psd}
\end{figure}

\begin{figure}[htb]
    \centering
    \includegraphics[width=0.5\textwidth]{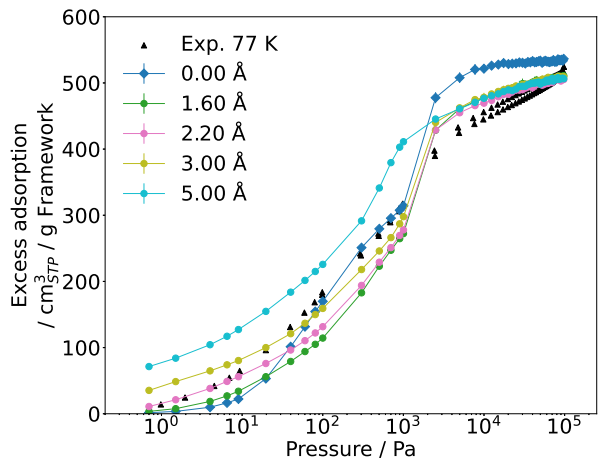}
    \caption{Pure component adsorption of nitrogen at 77 K in different random path COF structures. The COF layers are shifted by 1.60 \AA \space up to 5 \AA. All isotherms have been scaled with the previously discussed factor of 0.74. Lines are displayed to guide the eye.}
    \label{fgr:layer_shifting_N2_77}
\end{figure}

The layers of 2D COFs interact mainly by $\pi-\pi$-interactions and Van der Waals (vdW) forces, stabilizing the system in order to form crystalline porous structures. An often discussed feature of COFs is interlayer slipping, leading to larger lateral displacements between subsequent layers\textcolor{red}{\cite{Lukose2011}}. As a result, subsequent layers are no longer stacked directly on top of each other (eclipsed stacking motif) but are noticeably displaced to each other (shifted stacking motif). Therefore we study the effects of random displacements up to a maximum of 5~\AA \space generated using the random path generator and the procedure explained in section Random Path Generation. \textcolor{black}{In general the \textcolor{black}{slipping} is not necessarily random but can also be directed in a single direction. We do not consider this case in the present work. \textcolor{black}{Note that we used a constant interlayer distance of 3.601~\AA~for all random path structures, and also for the eclipsed structure in this part of the study}. Therefore the unit cell of the eclipsed structure becomes significantly smaller, which results in higher framework densities and thus lower loadings in comparison to an interlayer distance of 4.0505~\AA\textcolor{black}{~as obtained from DFT for the eclipsed case.}} For each hypothetical structure adsorption isotherms of argon, nitrogen and carbon dioxide, cumulative pore volume curves, enthalpy of adsorption at 0.98~Pa for argon, specific surface area and the void fraction are computed. This allows a systematic study of the effect of random interlayer slipping on those properties and finally on their effect on the adsorption isotherm.
\begin{figure}[ht]
    \centering
    \includegraphics[width=0.5\textwidth]{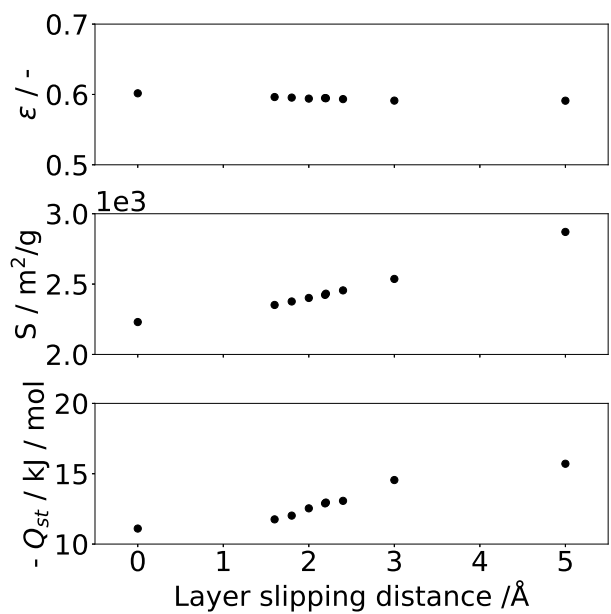}
    \caption{Effect of layer shifting distances on void fraction (calculated according to eq. (\ref{eqn:void_kb})), specific surface area and enthalpy of adsorption for argon at 0.98~Pa.}
    \label{fgr:shifting_mat_props}
\end{figure}
The cumulative pore volume curves plotted in Figure \ref{fgr:prob_dist} give a first impression on the geometric effects of interlayer slipping. The intersect of the cumulative pore volume curve and the axis of ordinate is the void fraction of each structure, which is only influenced slightly by the shifting. Small shifting distances lead to \textcolor{black}{steps in the curves}, indicative of more uniform pores. With increasing shifting distance the curves become more flat, indicating the pore size distribution is more heterogeneous. Especially the shifting of 5~\AA~results in areas in between three layers that interact as \textcolor{black}{nanocavities} which impact the adsorption behavior. 
\begin{figure}[ht!]
    \centering
    \includegraphics[width=0.9\textwidth]{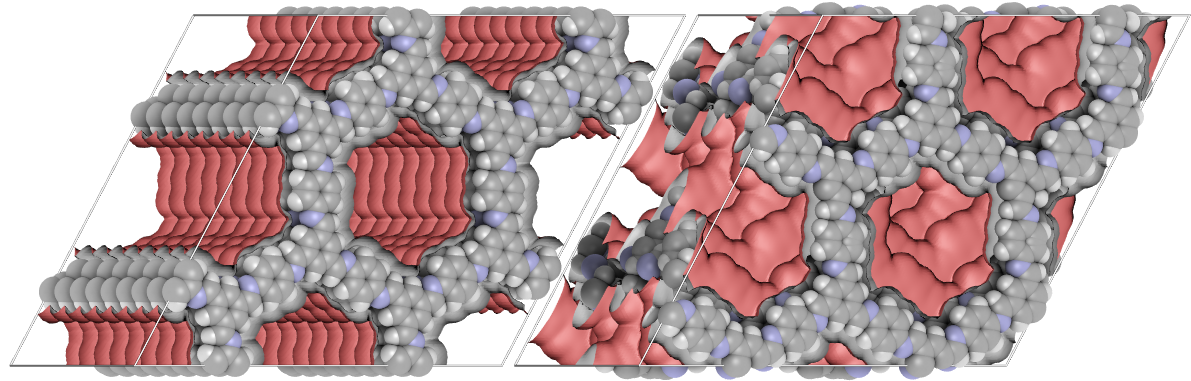}
\caption{\textcolor{black}{Visualisations of simulation boxes. Left: Eclipsed structure. Right: Random path structure with a \textcolor{black}{layer} \textcolor{black}{slipping} of 5~\AA. The increasing disorder and available surface becomes clearly visible. The random path structure also provides a visualization of "small" cavities in between three adjacent layers. Additional pictures are provided in the Supporting Information.}}
\label{fig.isosurface}
\end{figure}
Therefore the pore size distribution shown in Figure \ref{fgr:psd} becomes broader due to the different pore sizes found in the strongly shifted structure. The void fraction can also be determined by \textcolor{black}{random insertions} of helium. The results are plotted in Figure \ref{fgr:shifting_mat_props}. Small deviation \textcolor{black}{between the results obtained by \textcolor{black}{random insertions} and the pore volume curves} are due to differences in the methodology. It can be concluded that the free volume of the simulation box is almost independent of the interlayer \textcolor{black}{slipping} distance. \textcolor{black}{It can also be concluded that with increasing \textcolor{black}{slipping} distance the pore size is reduced by layers intruding \textcolor{black}{into} the main pores as illustrated in Fig.~\ref{fig.isosurface}.}
Another geometric property is the specific accessible surface area. 
With increasing \textcolor{black}{interlayer slipping} also the specific surface area increases as the surface of the layers intruding into the main pore becomes accessible since it is not covered by adjacent layers anymore.\\
\begin{figure}[htb]
    \centering
    \includegraphics[width=0.5\textwidth]{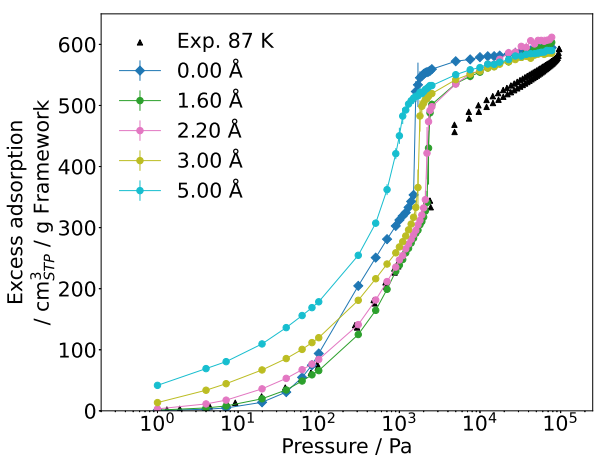}
    \caption{Pure component adsorption of argon at 87.3 K in different random path COF structures. The COF layers are shifted by 1.60 \AA \space up to 5 \AA. All isotherms have been scaled with the previously discussed factor of 0.74. Lines are displayed to guide the eye.}
    \label{fgr:layer_shifting_ar_87}
\end{figure}
Interestingly, a larger layer \textcolor{black}{slipping} introduces indentations between layers which appear to be highly favorable adsorption sites and result in high values of enthalpy of adsorption, computed at 0.98~Pa. At low pressures the enthalpy of adsorption only depends on the solid-fluid interactions and the framework geometry. The value increases with layer \textcolor{black}{slipping}, as seen in Figure \ref{fgr:shifting_mat_props}. \citeauthor{Forst2006}\cite{Forst2006} studied adsorption in chemically similar isoreticular MOFs with different pore sizes, which are a consequence of increasing linker sizes connecting the vertexes that contain the metal or metal oxide. It was found that adsorption isotherms can be divided into three regimes, with distinct shape effects determining the adsorption behavior. The first one at the lowest pressures is governed by the enthalpy of adsorption. For argon adsorption at 87 K (figure \ref{fgr:layer_shifting_ar_87}) and nitrogen adsorption at 77 K (Figure \ref{fgr:layer_shifting_N2_77}) in COF-LZU1 we found this regime at pressures below 100~Pa. It is followed by a pressure-regime where surface area determines the amount adsorbed, which we found from 100~Pa < p < 2000~Pa, and finally at higher pressures the decisive factor is the free volume. Figure~\ref{fgr:layer_shifting_ar_87} and~\ref{fgr:layer_shifting_N2_77} show the isotherms of various layer \textcolor{black}{slipping} motifs for argon at 87~K and nitrogen at 77~K, respectively. Results for both systems are equal so we discuss them together. The isotherms with a layer \textcolor{black}{slipping} up to 3 \AA \space show \textcolor{black}{capillary condensation} at $p \approx 2000\text{~Pa}$ whereas the isotherm of the structure shifted by 5 \AA \space does not show this behavior but a smoother transition to higher loadings. The smooth adsorption isotherms can be explained with the increasing disorder caused by the larger layer \textcolor{black}{slipping}. Therefore the indentations act like nanocavities with a diameter $\approx 3.8$~\AA, \space\textcolor{black}{which is two times the interlayer distance (3.6 \AA) minus the Lennard–Jones $\sigma$-parameter of carbon (3.47\AA)}. With those cavities filled with the adsorbates the pore diameter of the larger pores decreases for further adsorption and so \textcolor{black}{the cavities} show characteristic adsorption of smaller pores than reported for COF-LZU1. \textcolor{black}{\textcolor{black}{The resulting adsorption  behavior} is assumed to be similar for any 2D COF}. We can also emphasize that the isotherm of the eclipsed structure yields lower loadings for the low-pressure regime where the enthalpy of adsorption is the decisive factor. At $\approx$ 100~Pa, where the surface area determines the loading, adsorption is higher for the eclipsed structure than for the shifted structures. Close to saturation pressure where pores are filled, the eclipsed structure has the highest loading which matches its highest free volume. All \textcolor{black}{slipped} structures result in almost the same saturation loading, as expected because of similar void fractions. Concerning adsorption of nitrogen and argon at low temperatures it can be summarized that the DFT-optimized structure with an interlayer \textcolor{black}{slipping} of $\approx$ 2.19 \AA\ and the application of a scaling factor yields the best results in comparison with experiments.

\begin{figure}[hbt]
    \centering
    \includegraphics[width=0.5\textwidth]{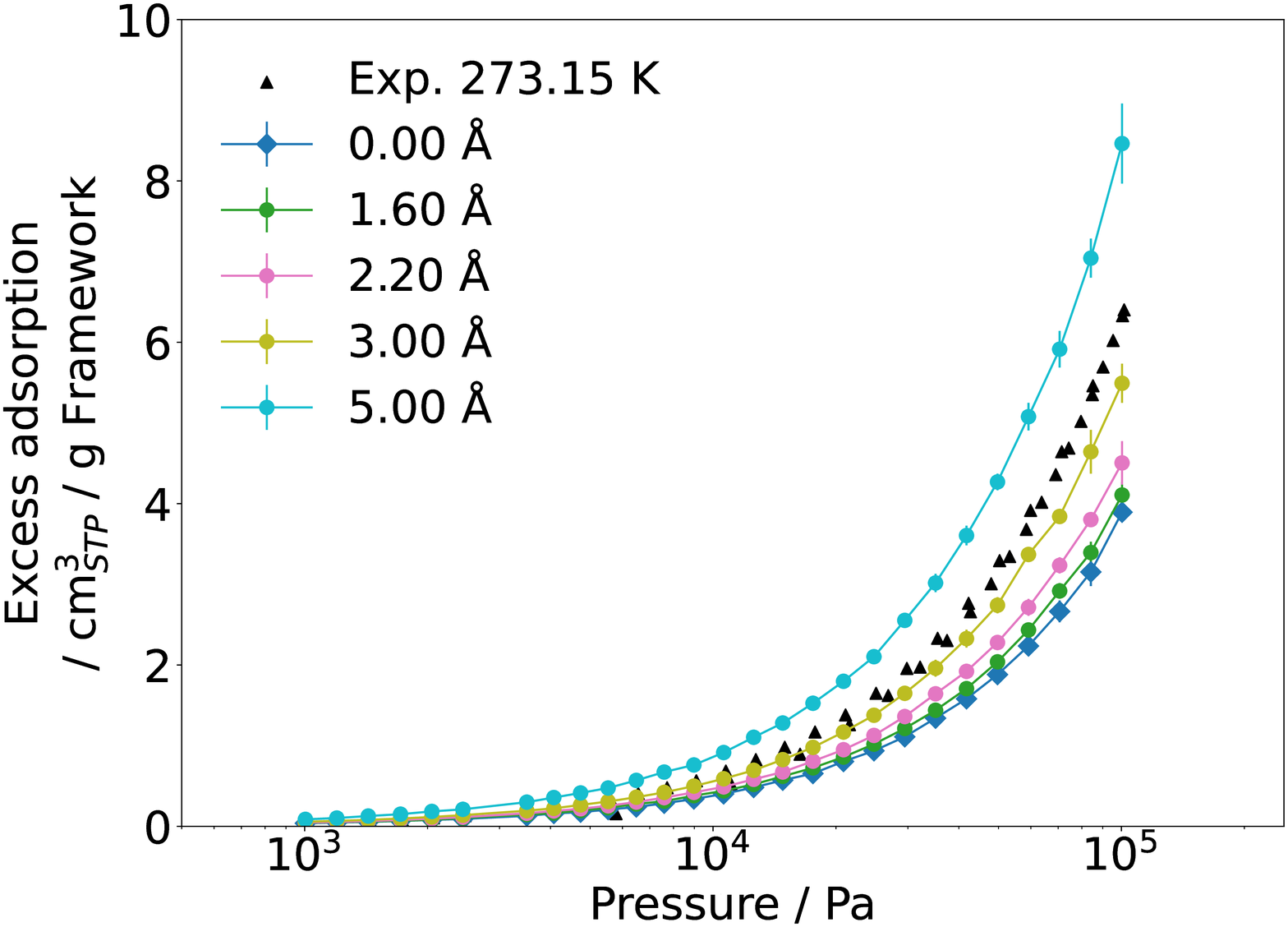}
    \caption{Pure component adsorption of nitrogen at 273.15 K in different random path COF structures. The COF layers are shifted by 1.60 \AA \space up to 5 \AA. Lines are displayed to guide the eye.}
    \label{fgr:layer_shifting_N2_273}
\end{figure}

\begin{figure}[hbt]
    \centering
    \includegraphics[width=0.5\textwidth]{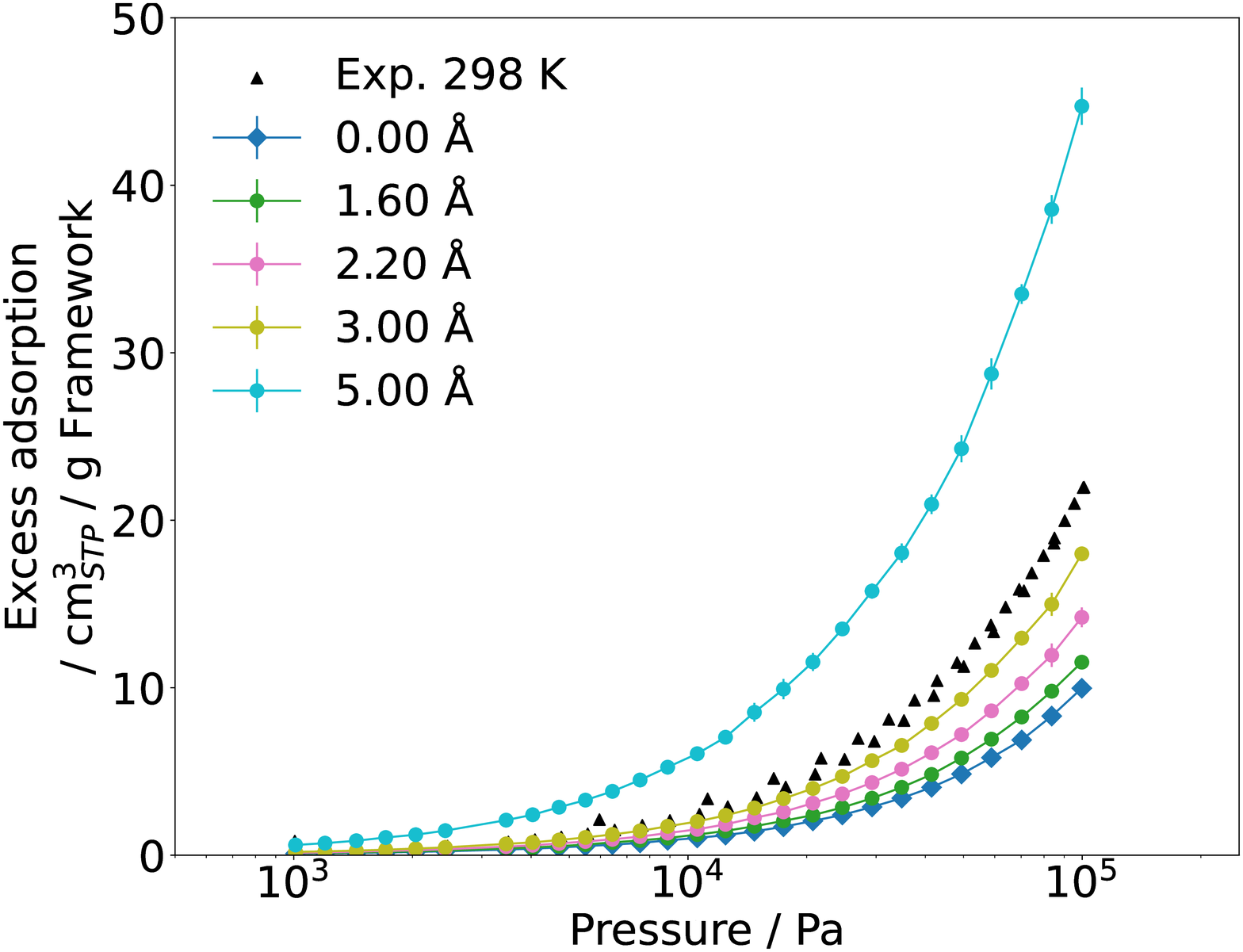}
    \caption{Pure component adsorption of carbon dioxide at 298 K in different random path COF structures. The COF layers are shifted by 1.60 \AA \space up to 5 \AA. Lines are displayed to guide the eye.}
    \label{fgr:layer_shifting_CO2_298}
\end{figure}

We carried out the same analysis for nitrogen at 273.15~K and for carbon dioxide at 298~K, see Figs~\ref{fgr:layer_shifting_N2_273} and \ref{fgr:layer_shifting_CO2_298}. For both isotherms no scaling has been applied. For both adsorbates the bulk states are supercritical and thus the loading in the frameworks is relatively low. Therefore adsorption will predominantly occur in the lower pressure regime (determined by enthalpy of adsorption). At those temperatures the DFT-optimized structure does not match well. For both cases a layer \textcolor{black}{slipping} of approx 3~\AA\, \textcolor{black}{results in loadings in good agreement to} experimental results. \textcolor{black}{We note that COF structures may thermally expand with increasing temperature - an effect we do not capture with our modelling approach with fixed frameworks. The impact on adsorption may not be negligible and should be addressed in a future study.}
\FloatBarrier

\section{Conclusion}
The present work aims at \textcolor{black}{clarifying the role of} framework geometry (especially) interlayer \textcolor{black}{slipping} on adsorption \textcolor{black}{of gases (argon, N$_2$ and CO$_2$)} in COF-LZU1. \textcolor{black}{In agreement to earlier studies we observe an} overestimation of the amount adsorbed in simulations relative to experiment. \textcolor{black}{The overestimated adsorption can not be explained by uncertainties in the molecular force field, suggesting there are defects or inclusions in the COF structure. More specifically} we found that the influence of framework point charges \textcolor{black}{for the considered COF} is of minor importance for adsorption of N$_2$ and CO$_2$. The Lennard–Jones energy parameters of the framework force field have a higher impact on the adsorption isotherm. However, possible deficiencies in the commonly used Dreiding force field are not expected to exclusively account for the overestimated adsorption. Following earlier work, a scaling factor for the simulated isotherms was introduced that relates the BET surface area of the model structure to that of the experiment. The saturation loading of the scaled isotherms is in good agreement with the experimental data. \textcolor{black}{Although there is no sophisticated method to measure imperfections in the COF framework directly, there \textcolor{black}{is much evidence pointing to} their existence\cite{Nguyen2016, Putz_2020,Alahakoon2020,Zeng2016}.}

The influence of interlayer \textcolor{black}{slipping} on various properties such as pore size distribution, enthalpy of adsorption, void fraction, accessible surface area and the amount adsorbed was studied by using hypothetical COF structures with different layer \textcolor{black}{slipping} in random directions. We demonstrate that interlayer slipping has a significant effect on the adsorption isotherm in the low-pressure regime, whereby layer slipping leads to an increased amount adsorbed. \textcolor{black}{For COF layers shifted randomly in lateral direction by $2 - 3$ \AA, we find good agreement between simulation results and experimental adsorption data.}
%
%
%
\textcolor{black}{This work shows \textcolor{black}{structure-property-relationships} that allow conclusions to be drawn from the adsorption isotherm \textcolor{black}{about} the framework \textcolor{black}{ structure} and its possible deviations from an ideal structure.}
\begin{acknowledgement}
This work was funded by the Deutsche Forschungsgemeinschaft (DFG, German Research Foundation) - Project-ID 358283783 - SFB 1333. 
We also thank the DFG for supporting this work by funding - EXC 2075/1 – 390740016 and EXC 2089/1-390776260 - under Germany's Excellence Strategy. We acknowledge the support by the Stuttgart Center for Simulation Science (SimTech).
Monte Carlo simulations were performed on the computational
resource BinAC at High Performance and Cloud Computing Group at the Zentrum f\"ur Datenverarbeitung of the University
of T\"ubingen, funded by the state of Baden-W\"urttemberg through bwHPC and the German Research Foundation (DFG) through grant no INST 37/935-1 FUGG.
\end{acknowledgement}

\begin{suppinfo}

Further details on the force field parameters, additional simulation results, CIF files of the COF structures, input files for RASPA and a Jupyter notebook containing the data analysis are found in the supporting material.

\end{suppinfo}

\appendix

\bibliography{achemso-demo}

\end{document}